\documentclass[pra,twocolumn,floatfix,a4paper,superscriptaddress]{revtex4}

\usepackage{bm,color,amsmath,txfonts}
\usepackage{graphicx}
\usepackage{siunitx}
\usepackage{subfigure}
\usepackage{verbatim}
\usepackage{dcolumn}
\usepackage{bm}
\usepackage{epsf}
\usepackage{xcolor}
\usepackage{hyperref}
\usepackage{hhline}
\usepackage{braket}
\usepackage{float}
\usepackage{enumerate}
\usepackage{bbm}
\usepackage{lipsum}

\begin{document}

\title{Proposal for optomagnonic teleportation and entanglement swapping}
\author{Zhi-Yuan Fan}
\affiliation{Interdisciplinary Center of Quantum Information, State Key Laboratory of Modern Optical Instrumentation, and Zhejiang Province Key Laboratory of Quantum Technology and Device, School of Physics, Zhejiang University, Hangzhou 310027, China}
\author{Xuan Zuo}
\affiliation{Interdisciplinary Center of Quantum Information, State Key Laboratory of Modern Optical Instrumentation, and Zhejiang Province Key Laboratory of Quantum Technology and Device, School of Physics, Zhejiang University, Hangzhou 310027, China}
\author{Hang Qian}
\affiliation{Interdisciplinary Center of Quantum Information, State Key Laboratory of Modern Optical Instrumentation, and Zhejiang Province Key Laboratory of Quantum Technology and Device, School of Physics, Zhejiang University, Hangzhou 310027, China}
\author{Jie Li}\thanks{jieli007@zju.edu.cn}
\affiliation{Interdisciplinary Center of Quantum Information, State Key Laboratory of Modern Optical Instrumentation, and Zhejiang Province Key Laboratory of Quantum Technology and Device, School of Physics, Zhejiang University, Hangzhou 310027, China}

\begin{abstract}	
A protocol for realizing discrete-variable quantum teleportation in an optomagnonic system is provided. Using optical pulses, an arbitrary photonic qubit state encoded in orthogonal polarizations is transferred onto the joint state of a pair of magnonic oscillators in two macroscopic yttrium-iron-garnet (YIG) spheres that are placed in an optical interferometer.  We further show that optomagnonic entanglement swapping can be realized in an extended dual-interferometer configuration with a joint Bell-state detection. Consequently, magnon Bell states are prepared. We analyze the effect of the residual thermal occupation of the magnon modes on the fidelity in both the teleportation and entanglement swapping protocols. 
\end{abstract}	

\date{\today}
\maketitle

\section{Introduction}

As an indispensable building block of quantum information science, quantum teleportation refers to the process of transferring an unknown quantum state at one location onto another quantum system some distance away. Ever since it was first proposed by Bennett {\it et al.} \cite{tele98}, quantum teleportation has been successfully realized in various physical systems over the past few decades. These include photons \cite{optical1,optical2,optical3,optical4}, nuclear spins \cite{NMR}, trapped ions \cite{ions1,ions2}, atomic ensembles \cite{ensemble1,ensemble2}, solid-state systems \cite{qubit}, high-frequency vibration phonons \cite{vibration}, and optomechanical systems \cite{optomechanics}, among many others. 
These successful demonstrations lay the foundation for realizing many other quantum protocols, such as quantum repeaters~\cite{repeat}, fault-tolerant quantum computation~\cite{computation}, etc.

Here we provide an optomagnonic quantum teleportation protocol which can transfer an arbitrary photonic qubit state to a dual-rail encoding magnonic system of two yttrium-iron-garnet (YIG) spheres.  We adopt an optical interferometer configuration, of which each arm contains a YIG sphere supporting an optomagnonic system. We use the Stokes type scattering of the magnon-induced Brillouin light scattering (BLS) to create an optomagnonic EPR state. A subsequent Bell-state measurement onto the input photonic qubit and the output Stokes photon from the interferometer enables such a photon-to-magnon quantum state transfer. The magnon state can be read out by activating the anti-Stokes scattering realizing the optomagnonic state-swap operation, from which the magnonic qubit state is retrieved onto the anti-Stokes photon.  We further propose an optomagnonic entanglement swapping protocol based on an extended dual-interferometer configuration, which realizes the transfer of the optomagnonic entanglement to a dual-rail encoding two-qubit magnonic system involving four YIG spheres and thus prepares a magnonic Bell state. 

In what follows, we first introduce the basic optomagnonic interactions in Section \ref{interaction}, and describe our two protocols respectively in Section \ref{telep} and Section \ref{swap}. We then analyze the effect of the magnon thermal excitations on the fidelity of the two protocols in Section \ref{thermal} and finally conclude in Section \ref{conc}.

\section{Optomagnonic interaction} \label{interaction}

In the past decade, hybrid magnonic systems based on collective spin excitations in ferrimagnetic materials, such as YIG, have gained significant attention due to their excellent ability to coherently couple with a variety of physical systems, including microwave photons \cite{mw1,mw2,mw3}, optical photons \cite{optomag1,optomag2,optomag3,optomag4,optomag5}, vibration phonons \cite{phonon1,phonon2,phonon3,phonon4,phonon5} and superconducting qubits \cite{supqu1,supqu2,supqu3,supqu4,supqu5}. The emerging field of hybrid quantum magnonics provides a platform not only for studying strong interactions between light and matter, but also for developing novel quantum technologies to be applied in quantum information processing, quantum sensing and quantum networks \cite{net1,net2,net3,prxQ}.  In particular, the coupling between magnons and optical photons, namely the optomagnonic interaction \cite{optomag1,optomag2,optomag3,optomag4,optomag5}, is indispensable for building a magnonic quantum network~\cite{prxQ}, where the transmission of the information between remote quantum nodes is realized by optical photons.  Such an optomagnonic interaction has been exploited in many proposals to cool magnons \cite{magcool}, prepare magnon Fock \cite{magfock}, cat \cite{magcat}, path-entangled \cite{WJW} states and magnon-photon entangled states \cite{Xie1,prxQ}, and realize magnon laser \cite{laser1,laser2}, frequency combs \cite{frecomb}, photon blockade \cite{Xie2}, polarization-state engineering \cite{Wuy}, etc.


\begin{figure*}[t]
\centering
\hskip-0.5cm\includegraphics[width=0.75\linewidth]{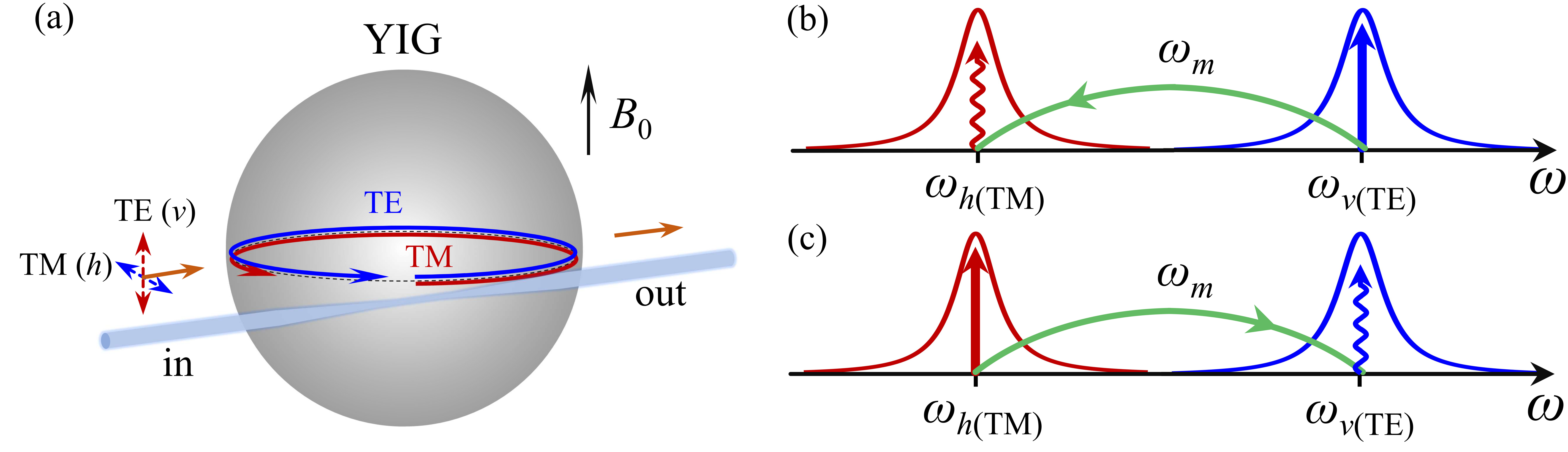} 
	\caption{(a) A typical optomagnonic system: a YIG sphere supports a magnon mode and two optical WGMs. (b) Mode frequencies corresponding to the optomagnonic Stokes scattering, which leads to the PDC interaction. (c) Mode frequencies associated with the optomagnonic anti-Stokes scattering, which yields the state-swap (BS) interaction.}
	\label{fig1}
\end{figure*}

The typical optomagnomic system, as depicted in Figure \ref{fig1}(a), consists of a YIG sphere, which supports both a magnetostatic mode (i.e., the magnon mode) and optical whispering gallery modes (WGMs). The WGM resonator is near the surface of the YIG sphere and the input optical field is evanescently coupled to the WGM, e.g., via a tapered fiber \cite{optomag1,optomag2} or prism \cite{optomag3,optomag4}.  The optomagnonic interaction in such a system is embodied by the magnon-induced BLS, where the photons of a WGM are scattered by lower-frequency magnons, typically in GHz \cite{optomag1,optomag2,optomag3,optomag4}, giving rise to sideband photons with their frequency shifted by the magnon frequency and their polarization changed.  When the frequency of the scattered photons matches another WGM, namely the triple resonance condition, the optomagnonic scattering probability is maximized. This can be easily achieved by tuning the magnon frequency via changing the strength of the bias magnetic field ($B_0$).   Typically, there are two types of magnon-induced BLSs, i.e., the Stokes and anti-Stokes scatterings, corresponding to optomagnonic parametric down conversion (PDC) and state-swap (beam-splitter, BS) interactions, respectively \cite{prxQ,WJW}.  In the optomagnonic scattering, the angular momenta of the WGM photons and magnons are conserved, which leads to a selection rule and prominent asymmetry in the Stokes and anti-Stokes scattering strengths~\cite{SR1,SR2,SR3,SR4}. This allows us to select a particular type of the optomagnonic interactions (either PDC or BS) on demand.

The Hamiltonian accounting for the optomagnonic interaction in such a three-mode system reads
\begin{equation}
	\begin{split} \label{Hamiltonian}
		H/\hbar=&\ \omega_v a^\dagger a+\omega_h b^\dagger b+\omega_m m^\dagger m \\
		&+g_0(a^\dagger b m+a b^\dagger m^\dagger)+H_{\rm{dri}}/\hbar,
	\end{split}
\end{equation}
where $j\,\,{=}\,\,a,b$ ($j^\dagger$) and $m$ ($m^\dagger$) are the annihilation (creation) operators of the two WGMs and the magnon mode, respectively, and $\omega_k$ ($k\,\,{=}\,\,v,h,m$) correspond to their resonance frequencies, satisfying the relations $|\omega_v - \omega_h|=\omega_m$ and $\omega_m \ll \omega_{v,h}$. Here the subscripts $v$ and $h$ represent two orthogonal polarizations of the two WGMs, i.e., the transverse-electric (TE) mode and the transverse-magnetic (TM) mode. The optomagnonic interaction is a three-wave process and $g_0$ is the corresponding single-photon coupling strength. The last term denotes the driving Hamiltonian $H_{\rm{dri}}=i\hbar E_j (j^\dagger e^{-i\omega_0 t}-j e^{i\omega_0 t})$, where $E_j=\sqrt{\kappa_j P_j/(\hbar \omega_0)}$ is the coupling strength between the WGM $j$ (with decay rate $\kappa_j$) and the optical drive field with frequency $\omega_0$ and power $P_j$.  To enhance the optomagnonic coupling strength, a strong drive field is used to resonantly pump one of the WGMs, i.e., $\omega_0 = \omega_v$ or $\omega_h$, depending on which type of the optomagnonic interactions (either PDC or BS) is desired. 

For the case where the TE WGM ($a$) is pumped, i.e., $\omega_0=\omega_v$ and $\Delta_b \equiv \omega_h-\omega_0=-\omega_m$, cf. Figure \ref{fig1}(b), the strongly driven WGM $a$ can be treated classically as a number $\alpha \equiv \langle a \rangle$ \cite{prxQ}, and the effective optomagnonic Hamiltonian then becomes the PDC form, $H_{\rm{int}}^{S} = \hbar g(b m+b^\dagger m^\dagger)$, with $g=g_0 \alpha$ being the effective optomagnonic coupling strength. This interaction corresponds to the Stokes scattering, where the TE-WGM photons convert into lower-frequency sideband photons (resonant with the TM WGM $b$) by creating magnon excitations. Such a PDC interaction can be used to generate optomagnonic entangled states \cite{Xie1,prxQ}. 
Specifically, the WGM $b$ and magnon mode $m$ are prepared in a two-mode squeezed state (unnormalized)
\begin{equation}\label{eqTMS}
| \psi \rangle_{b,m}= |00 \rangle_{b,m} + \!\! \sqrt{P} |11 \rangle_{b,m} + {\cal O}(P) \,\, ,
\end{equation}
where $P$ is the probability for a single Stokes scattering event to occur and ${\cal O}(P)$ denotes the higher-excitation terms, of which the probabilities are equal to or smaller than $P^2$. The scattering probability increases with the power of the drive field, and when the power is sufficiently weak, the scattering probability $P \ll 1$. In this weak-coupling limit, the probability of generating two-magnon (photon) state $|2 \rangle_{m(b)}$ and higher-excitation states $|n \rangle_{m(b)}$ ($n\,\,{>}\,\,2$) is negligibly small. Such a low scattering probability of creating an entangled pair of single excitations was adopted in Ref.~\cite{WJW}, which suggests an optomagnonic variant of the Duan-Lukin-Cirac-Zoller protocol \cite{DLCZ}. It was also used in cavity optomechanical experiments for creating entangled states of single photons and phonons~\cite{Simon16,Simon18}.
 
Similarly, when the TM WGM ($b$) is resonantly pumped, i.e., $\omega_0=\omega_h$ and $\Delta_a \equiv \omega_v-\omega_0=\omega_m$, cf. Figure \ref{fig1}(c), the anti-Stokes scattering is activated and the effective optomagnonic Hamiltonian becomes the BS type, $H_{\rm{int}}^{AS}=\hbar g(a^\dagger m+a m^\dagger)$, with the effective coupling $g=g_0 \beta$ and $\beta \equiv \langle b \rangle$. This interaction realizes the state-swap operation between the magnon mode and the WGM $a$, accompanied with the process that TM-WGM photons convert into higher-frequency anti-Stokes photons (resonant with the TE WGM $a$) by eliminating magnon excitations. As will be shown later, this interaction is used to read out the magnon state.


\begin{figure*}[t]
\centering
\hskip-1cm\includegraphics[width=0.8\linewidth]{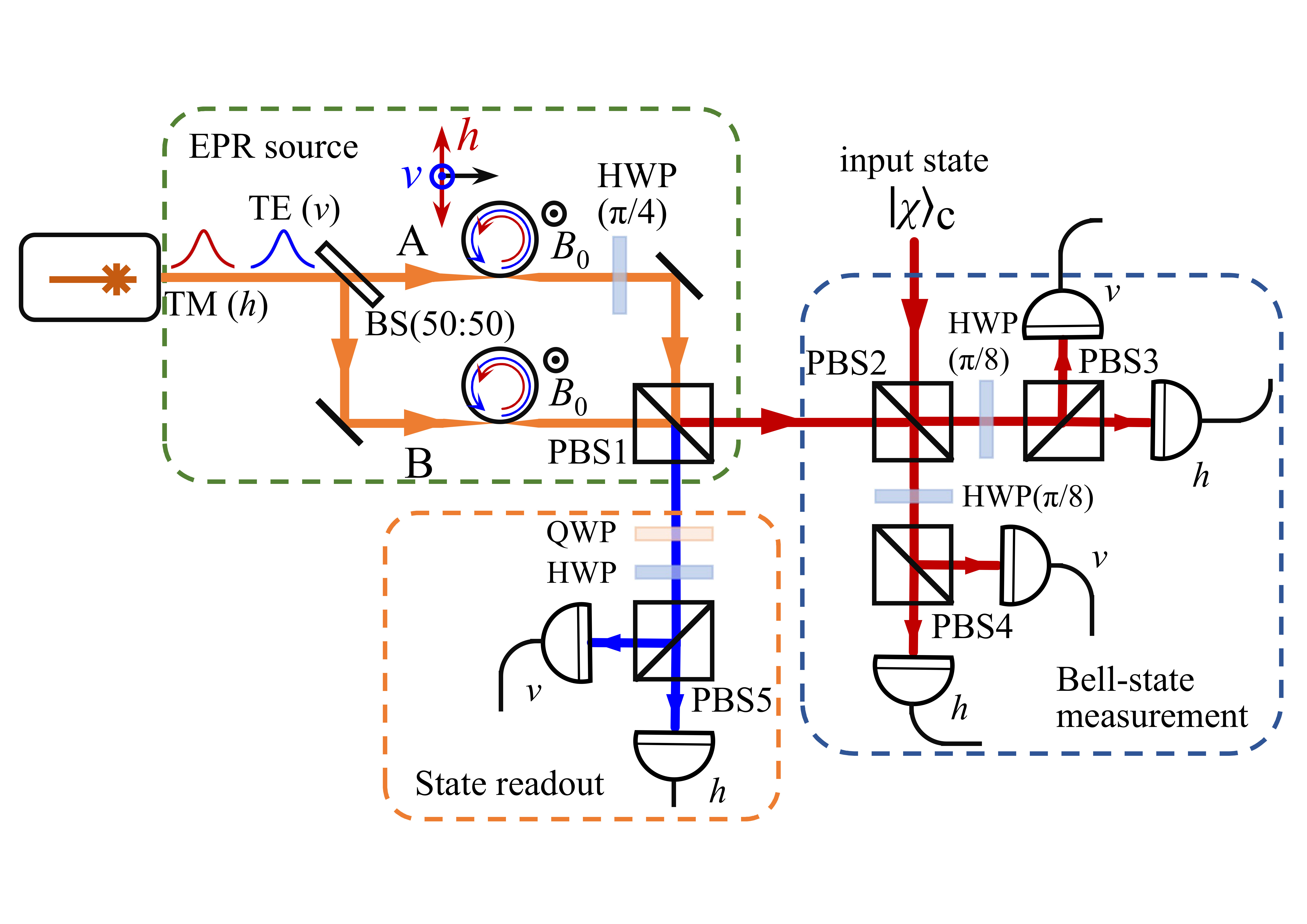} 
	\caption{Sketch of the optomagnonic quantum teleportation protocol. It  consists of three steps: preparation of the optomagnonic EPR state, Bell-state measurement and readout of the magnonic state. In the EPR-state preparation setup, an optical interferometer configuration is adopted and its each arm contains a YIG sphere which supports an optomagnonic system of a magnon mode and two optical WGMs. See text for detailed descriptions. }
\label{fig2}
\end{figure*}

\section{Optomagnonic quantum teleportation}\label{telep}


We now proceed to describe our optomagnonic teleportation protocol, which is able to transfer a photonic qubit state (in polarization encoding) onto a magnonic system consisting of two optomagnonic devices placed in two arms of an optical interferometer, cf. Figure \ref{fig2}. The two magnon oscillators are subject to a simultaneous excitation using a {\it weak} pulse that drives the TE WGM to activate the optomagnonic Stokes scattering.  After a Bell-state measurement of the input photon with the Stokes photon from the optomagnonic devices, the input photonic qubit state is then transferred onto the dual-rail encoding magnonic system.  

For simplicity, we assume at most single excitations in the optomagnonic devices. This is the case of using a weak pulse, where the probability of creating higher-excitation states $|n \rangle_{m(b)}$ ($n\,\,{\ge}\,\,2$) in the Stokes scattering is negligible.  Since the vacuum component of the state \eqref{eqTMS} will not trigger any coincidences in the Bell-state detection, leading to unsuccessful trials for the teleportation, the protocol can be described using a simplified model \cite{Jie20}, where a TE-polarized single-photon pulse is sent onto a 50/50 BS to drive the optomagnonic devices in the interferometer. After the BS, an optical path-entangled state, $\frac{1}{\sqrt{2}}(\ket{01}_{AB}+\ket{10}_{AB})$, is generated in the two outputs (i.e., the upper path A and lower path B). In each path, the pulse resonantly drives the TE WGM of the YIG sphere to activate the Stokes scattering.  By selecting trials with successful scattering events, the PDC interaction prepares an optomagnonic Bell state in the form of
\begin{equation}\label{EPR}
	\ket{\phi^+}_{bm}= \frac{1}{\sqrt{2}} \left( \ket{H}_b\ket{L}_{m}+\ket{V}_b\ket{U}_{m} \right),
\end{equation}
where $\ket{L}_m$ ($\ket{U}_m$) denotes the generated single magnon is at the lower path B (the upper path A), and  $\ket{H}_b$ ( $\ket{V}_b$) represents the accompanied Stokes photon is in the horizontal (vertical) polarization. The Stokes photon, with equal probability in path A or B, then couples to the nanofiber or the prism coupler and enters the first polarizing beam splitter (PBS1) before going to the next stage of the Bell-state detection.  Note that the polarization of the Stokes photon in the upper path is changed from $H$ to $V$ after passing through a half-wave plate (HWP).   We remark that the unsuccessful scattering events leave the single photons remaining in the vertical polarization. These photons eventually enter the other output of the PBS1 and thus have no impact on the subsequent Bell-state measurement.  We also assume that the magnon modes are initially in their quantum ground state, which is the case for GHz magnons at a low temperature, e.g., of 10 mK. Nevertheless, in Section \ref{thermal} we shall discuss the effect of the residual thermal occupation on the fidelity of the teleportation.

The input photonic qubit state to be teleported is an arbitrary superposition of two polarization modes, i.e., $\ket{\chi}_c=\alpha\ket{H}_c+\beta\ket{V}_c$, with the complex coefficients $\alpha$ and $\beta$ satisfying $|\alpha|^2+|\beta|^2=1$. Such an arbitrary qubit state can be constructed on the surface of the polarization Poincare sphere by using a HWP and a quarter-wave plate (QWP).  This input photon, being resonant with the TM WGM, goes into one input of the PBS2 and meanwhile the output of the PBS1 enters the other input port.  Thereby, the joint state before the Bell-state measurement is as follows:
\begin{equation} \label{overall}
	\begin{split}
		\ket{\phi^+}_{bm}\otimes \ket{\chi}_c  \! &=\! \frac{1}{\sqrt{2}} \big( \alpha \ket{H}_b\ket{H}_c\ket{L}_m + \beta\ket{H}_b\ket{V}_c\ket{L}_m \\
		&\quad\ +\alpha \ket{V}_b\ket{H}_c \ket{U}_m+\beta \ket{V}_b \ket{V}_c\ket{U}_m \big).
	\end{split}
\end{equation}
The Bell-state measurement is performed onto the input photonic qubit ($c$) and the output Stokes photon ($b$) from the interferometer, and the detection setup consists of a HWP, a PBS and two single-photon detectors in each output of the PBS2, cf. Figure \ref{fig2}. A coincidence measurement  projects the optical modes onto the polarized Bell states $\ket{\phi^{\pm}}_{bc}=\frac{1}{\sqrt{2}}(\ket{H}_b\ket{H}_c\pm\ket{V}_b\ket{V}_c)$.  The fast axis of the HWP is set at $22.5^{\circ}$, which acts as a Hadamard operation on the polarization of the photons passing through PBS2.  The Bell states $\ket{\phi^{\pm}}_{bc}$ correspond to different coincidence measurements, as can be seen from the following
\begin{equation}
	\begin{split}
		\ket{\phi^{+}}_{bc}
		&\xrightarrow{\mathrm{PBS2 \,\&\, HWP}}\frac{1}{\sqrt{2}} \left(a_{3,h}^\dagger a_{4,h}^\dagger+a_{3,v}^\dagger a_{4,v}^\dagger\right)\ket{\rm{vac}}, \\
		\ket{\phi^{-}}_{bc}&\xrightarrow{\mathrm{PBS2 \,\&\, HWP}}\frac{1}{\sqrt{2}} \left(a_{3,h}^\dagger a_{4,v}^\dagger+a_{3,v}^\dagger a_{4,h}^\dagger\right)\ket{\rm{vac}},
	\end{split}
\end{equation}
where the operator $a_{i,h}^\dagger$ ($a_{i,v}^\dagger$) denotes the detection of a single horizontally (vertically)-polarized photon at one of the two outputs of the PBS$i$ ($i=3,4$), and $\ket{\rm{vac}}$ means the vacuum state. 
Note that, in addition to the Bell states $\ket{\phi^{\pm}}_{bc}$, the other two types of Bell states $\ket{\psi^{\pm}}_{bc}=\frac{1}{\sqrt{2}}(\ket{H}_b\ket{V}_c\pm\ket{V}_b\ket{H}_c)$ can be realized by using photon-number-resolving detectors, and we disregard these cases for simplicity.

The measurement that projects the optical modes $b$ and $c$ onto the Bell state $\ket{\phi^+}_{bc}$ projects the magnonic system onto the state
\begin{equation}
	\ket{\chi'}_m=\alpha \ket{L}_m+\beta \ket{U}_m,
\end{equation}
which indicates the successful teleportation of the input photonic qubit state $\ket{\chi}_c$ to a dual-rail encoding magnonic system, and corresponds to the ideal quantum teleportation without requiring additional correction operations in the readout step.  On the other hand, the measurement associated with the Bell state $\ket{\phi^-}_{bc}$ projects the magnonic system onto the state 
 \begin{equation}
 	 \ket{\chi''}_m=\alpha \ket{L}_m-\beta \ket{U}_m.
 \end{equation}
 It has a $\pi$-phase difference with respect to the input optical state $\ket{\chi}_c$, which can be corrected in the readout step using a feed-forward operation by applying a phase shift in the optical interferometer \cite{Jie20,optomechanics}.

To verify the successful teleportation, we need to retrieve the teleported magnonic qubit state. To achieve this, we exploit the optomagnonic state-swap (BS) interaction, as introduced in Section \ref{interaction}, where a TM-polarized weak pulse is sent to the interferometer to activate the anti-Stokes scattering and the magnonic state is then transferred to the anti-Stokes photon. Due to the orthogonal polarization with respect to the Stokes photon (produced in the Stokes scattering), the anti-Stokes photon leaves from the other output of the PBS1, and enters the state-readout setup (cf. Figure \ref{fig2}) \cite{optomechanics}, from which the magnonic qubit state is retrieved.

\begin{figure*}[t]
\centering
\hskip-0.5cm\includegraphics[width=0.85\linewidth]{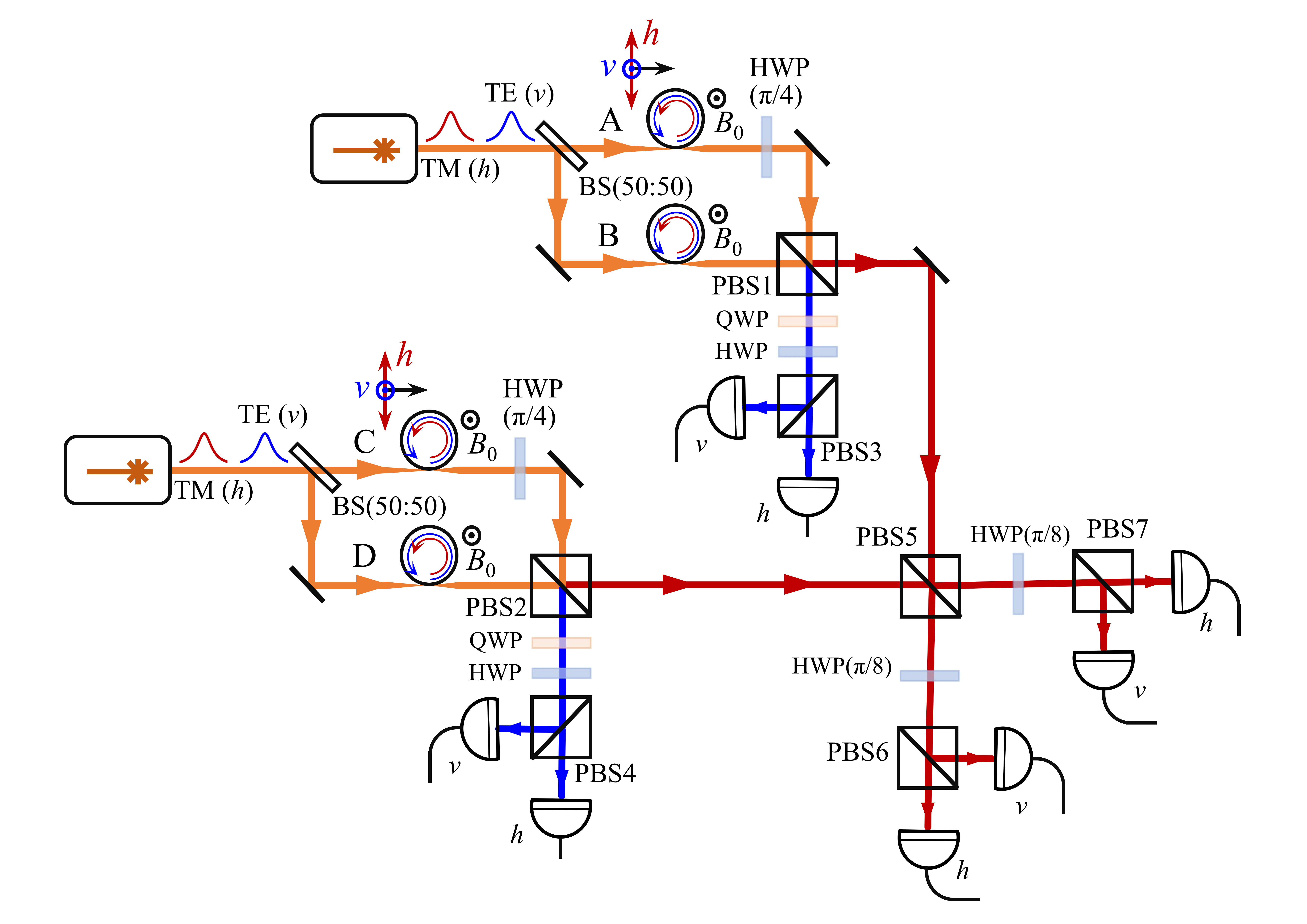} 
	\caption{Sketch of the optomagnonic entanglement swapping protocol. It is based on a dual-interferometer configuration, combined with a joint Bell-state detection and magnonic state readout devices. }
	\label{fig3}
\end{figure*}

\section{Optomagnonic entanglement swapping}\label{swap}

Quantum entanglement plays an essential role in all the quantum-teleportation related protocols. 
Naturally, entangled states can be obtained in directly coupled systems, such as the EPR state produced in the optomagnonic Stokes scattering.  For two systems that have no direct interaction, quantum mechanics also manifests its unique capabilities to establish quantum entanglement between them, one of which is referred to the entanglement swapping \cite{tele98,optical2,enswap}. In this section, we show that the entanglement swapping protocol allows us to prepare the magnon modes in space-separated YIG spheres into an entangled Bell state.   

The detailed entanglement swapping protocol is shown in Figure \ref{fig3}, which consists of two optical interferometer setups used in Section \ref{telep}, a joint Bell-state detection and the associated state-readout devices.  Similarly to the teleportation protocol, in each interferometer setup containing two YIG spheres, a TE-polarized single-photon pulse is sent to activate the Stokes scattering, which prepares an optomagnonic Bell entangled state $\ket{\phi^+}_{bm}$, as in Equation \eqref{EPR}.  Therefore, the overall state of the two interferometer setups before performing a joint Bell-state measurement on the scattered Stokes photons reads
\begin{equation}\label{overall2}
	\begin{split}
		\ket{\Psi}_{\rm{total}}=&\ \ket{\phi^+}_{b_1m_1}\otimes\ket{\phi^+}_{b_2m_2} \\
		=&\ \frac{1}{2} \big( \ket{H}_{b_1}\ket{L}_{m_1}+\ket{V}_{b_1}\ket{U}_{m_1} \big)\\
		&\times \big( \ket{H}_{b_2}\ket{L}_{m_2}+\ket{V}_{b_2}\ket{U}_{m_2} \big),
	\end{split}
\end{equation}
where the subscripts $1$ and $2$ are used to distinguish the two interferometers, and the notation is the same as in Equation \eqref{EPR}.  The above state can be rewritten in the basis of Bell states as
\begin{equation}\label{4Bell}
	\begin{split}
		\ket{\Psi}_{\rm{total}}=&\ \frac{1}{2} \Big( \ket{\psi^+}_{b_1b_2}\ket{\psi^+}_{m_1m_2} + \ket{\psi^-}_{b_1b_2}\ket{\psi^-}_{m_1m_2} \\
		& +\ket{\phi^+}_{b_1b_2} \ket{\phi^+}_{m_1m_2} + \ket{\phi^-}_{b_1b_2} \ket{\phi^-}_{m_1m_2} \Big),
	\end{split}
\end{equation}
where $\ket{\phi^{\pm}}_{b_1b_2}$ and $\ket{\psi^{\pm}}_{b_1b_2}$ are the four Bell states of the Stokes photons in the two interferometers, which take the same form as $\ket{\phi^{\pm}}_{bc}$ and $\ket{\psi^{\pm}}_{bc}$ in Section \ref{telep}. $\ket{\phi^\pm}_{m_1m_2}$ and $\ket{\psi^\pm}_{m_1m_2}$ are the four Bell states of a pair of dual-rail encoding magnonic systems, defined as $\ket{\phi^\pm}_{m_1m_2} = \frac{1}{\sqrt{2}} ( \ket{L}_{m_1}\ket{L}_{m_2}\pm\ket{U}_{m_1}\ket{U}_{m_2})$ and $\ket{\psi^\pm}_{m_1m_2} = \frac{1}{\sqrt{2}} (\ket{L}_{m_1}\ket{U}_{m_2}\pm\ket{U}_{m_1}\ket{L}_{m_2})$.

From Equation \eqref{4Bell}, it is clear to see that a Bell-state measurement on the Stokes photons from the outputs of the two interferometers projects the magnonic systems onto a corresponding Bell state. Specifically, a coincidence measurement corresponding to the optical Bell state $\ket{\phi^+}_{b_1b_2}$ ($\ket{\phi^-}_{b_1b_2}$) projects the magnonic systems onto the Bell state $\ket{\phi^+}_{m_1m_2}$ ($\ket{\phi^-}_{m_1m_2}$).  This implies that the magnonic systems establish the same form of entanglement as the corresponding optical Bell state.  Similarly as in the teleportation protocol, we disregard the other two types of the Bell-state measurements associated with $\ket{\psi^{\pm}}_{b_1b_2}$.

The entanglement swapping can also be understood in the framework of quantum teleportation in the sense that it transfers an optomagnonic quantum correlation (instead of a photonic qubit state $\ket{\chi}_c$) to the magnonic systems. This can be seen by comparing Eqs. \eqref{overall} and \eqref{overall2}. This further reflects the versatility of the quantum teleportation protocol, which can transfer not only a qubit state but also multi-qubit states, e.g., quantum correlations. Replacing the optomagnonic system with an optomechanical system~\cite{Simon18} in one of the interferometers also allows us to prepare a hybrid magnon-phonon Bell state, which may find potential applications in hybrid quantum networks \cite{prxQ}.

 

\section{Effect of magnonic thermal excitations}\label{thermal}

\begin{figure}[t]
\hskip-0.2cm	\includegraphics[width=0.85\linewidth]{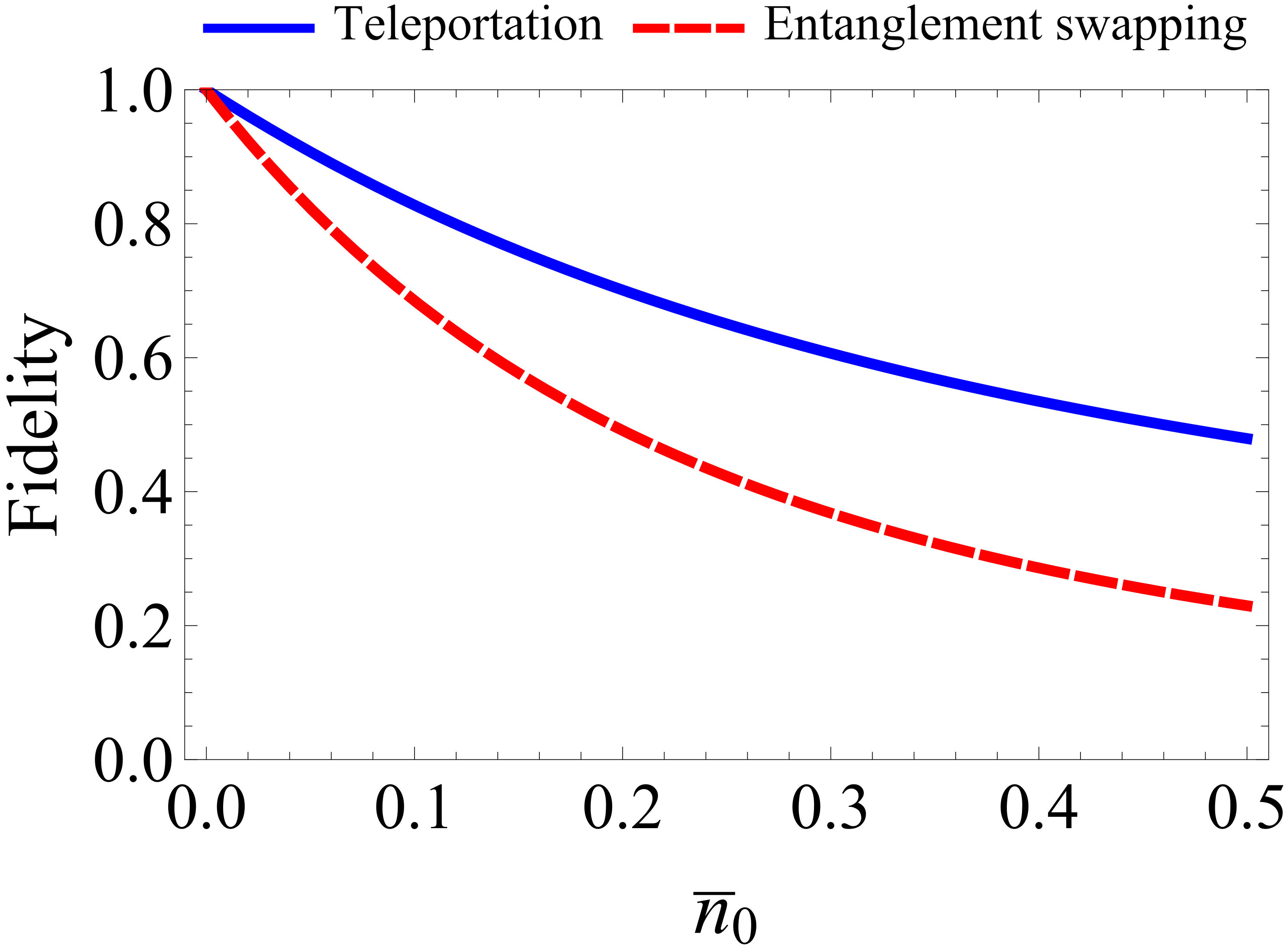} 
	\caption{Fidelity in the teleportation (solid) and entanglement swapping (dashed) protocol versus the thermal occupation $\bar{n}_0$ of the magnon modes.}
	\label{fig4}
\end{figure}

In the teleportation and entanglement swapping protocols, we neglect the dissipation of the magnon modes. This is the case of using fast optical pulses~\cite{Simon16} such that the magnon dissipation can be assumed to be negligible during an experimental run.  We also assume that the magnon modes are initially prepared in their quantum ground state. This is a good approximation for the magnon modes at $\sim$ GHz frequency \cite{optomag1,optomag2,optomag3,optomag4,optomag5} and at low temperature of, e.g., 10 mK. However, the optical pulses may heat the magnon modes due to the optical absorption of the YIG, causing the magnon modes to be at a thermal state. To include this heating effect in practical situations, we assume that the magnon modes are initially prepared in a thermal state $\rho_{th}=(1-s)\sum_{n=0}^{\infty} s^n\ket{n}\bra{n}$, with $s=\bar{n}_0/(\bar{n}_0+1)$ and a thermal occupation $\bar{n}_0\ll 1$. Note that the frequencies of the magnon modes are assumed to be (nearly) identical and thus they have equal thermal occupation, i.e., they are in the same thermal state. For a small $\bar{n}_0<0.2$, $s<0.17$, $s^2<0.03$ and $s^3<0.005$, and the total probability of higher-excitation terms $\ket{n}$ ($n>2$) is less than $0.5\%$. Thus we can safely approximate $\rho_{th}\simeq (1-s) \left(\ket{0}\bra{0}+s\ket{1}\bra{1}+s^2\ket{2}\bra{2} \right)$.   The density matrix of the two magnon modes (in path A and B) in the teleportation scheme is then
\begin{equation} \label{initial1}
	\rho_m=\rho_A\otimes\rho_B\simeq (1-s)^2\sum_{n_A,n_B=0}^2 s^{n_A+n_B}\ket{n_A n_B}\bra{n_A n_B},
\end{equation}
which is a probabilistic mixture of nine pure states $\ket{n_A n_B}$ ($n_A, n_B=0,1,2$). This mixed initial state eventually leads to the following teleported magnonic state after the Bell-state measurement associated with $\ket{\phi^+}_{bc}$:
\begin{equation}\label{rhom}
	\rho'_m=(1-s)^2\sum_{n_A,n_B=0}^2 s^{n_A+n_B} \ket{\varphi_{n_A n_B}}\bra{\varphi_{n_A n_B}},
\end{equation}
where $\ket{\varphi_{n_A n_B}}=\alpha\ket{n_A,n_B+1}_{AB}+\beta\ket{n_A+1,n_B}_{AB}$ is the teleported magnonic state corresponding to the pure inital state $\ket{n_A n_B}$ in Equation \eqref{initial1}. Clearly, $\ket{\varphi_{00}}=\alpha\ket{01}_{AB}+\beta\ket{10}_{AB} \equiv \alpha \ket{L}_m+\beta \ket{U}_m$ corresponds to the ideal initial state $\ket{00}_{AB}$ considered in Section \ref{telep}. For the Bell-state detection related to $\ket{\phi^-}_{bc}$, we obtain the same $\rho'_m$ as in Equation \eqref{rhom} but with $\beta$ replaced by $-\beta$ in $\ket{\varphi_{n_A n_B}}$.   All other states in $\rho'_m$ are orthogonal to the desired state $\ket{\varphi_{00}}$, and result in a reduction of the teleportation fidelity, which is
\begin{equation}\label{fff1}
	\mathcal{F}_1=\bra{\varphi_{00}}\rho'_m\ket{\varphi_{00}}=1/\big(1+s+s^2 \big)^2.
\end{equation}
The solid line in Figure \ref{fig4} clearly shows a declining fidelity versus the thermal occupation $\bar{n}_0$.  For a genuine quantum teleportation with the fidelity $\mathcal{F}_1 >2/3$ \cite{Pope}, a small $\bar{n}_0 \le \, \sim\,$0.2 is required. This is similar to the finding in the optomechanical teleportation \cite{Jie20}.

By contrast, in the entanglement swapping protocol, four magnon modes (in path A, B, C and D, respectively) are involved, cf. Figure \ref{fig3}. 
Following the same approach, we obtain the final joint state of the magnonic systems, after the Bell-state measurement associated with $\ket{\phi^{\pm}}_{b_1b_2}$, given by
\begin{equation}
	\rho'_{m_1 m_2}=(1-s)^4 \!\! \sum_{n_{A(B)}=0}^2 \sum_{n_{C(D)}=0}^2 s^{n_A+n_B} s^{n_C+n_D} \ket{\varphi_{m_1m_2}}\bra{\varphi_{m_1m_2}},
\end{equation}
where $A$, $B$, $C$ and $D$ are used to distinguish the magnon modes via their path information, and $\ket{\varphi_{m_1m_2}}=\frac{1}{\sqrt{2}}\left( \ket{n_A,n_B+1,n_C,n_D+1}_{m_1m_2} \pm \ket{n_A+1,n_B,n_C+1,n_D}_{m_1m_2} \right)$ is the teleported magnonic state corresponding to the pure state $\ket{n_A \,n_B\, n_C\, n_D}$ in the mixed initial state $\rho_{m_1m_2}=\rho_{m_1} \otimes \rho_{m_2}$, cf. Equation \eqref{initial1}. For the ideal case of the initial ground state considered in Section \ref{swap}, we obtain the desired states $\ket{\varphi_{m_1m_2}} =\frac{1}{\sqrt{2}} \left(\ket{0101}_{m_1m_2} \pm \ket{1010}_{m_1m_2} \right) \equiv \ket{\phi^{\pm}}_{m_1m_2}$ after the entanglement swapping. The other additional terms in $\rho'_{m_1 m_2}$ are related to the residual thermal excitations in the magnonic initial state, which are unwanted and reduce the fidelity in the entanglement swapping protocol.  The corresponding fidelity is given by
\begin{equation}\label{fff2}
	\mathcal{F}_2\,{=}\,_{m_2m_1}\!\!\bra{\phi^{\pm}}\rho'_{m_1m_2}\ket{\phi^{\pm}}_{m_1m_2}=1/\big(1+s+s^2\big)^4.
\end{equation}
The dashed line in Figure \ref{fig4} shows a decreasing fidelity $\mathcal{F}_2$ as the thermal occupation $\bar{n}_0$ increases. It reduces more rapidly with respect to the fidelity $\mathcal{F}_1$ in the teleportation protocol because of the relation $\mathcal{F}_2 = \mathcal{F}_1^2$, as seen from Equations \eqref{fff1} and \eqref{fff2}.


\section{Conclusion}\label{conc}

We present two protocols for realizing optomagnonic quantum teleportation and entanglement swapping, respectively, adopting YIG spheres and an optical interferometer configuration. The optomagnonic Stokes and anti-Stokes scatterings are the essential elements for preparing optomagnonic EPR states and optically reading out the magnonic states. A Bell-state detection enables the transfer of an arbitrary photonic qubit state to a dual-rail encoding magnonic system in the former protocol, and the transfer of the optomagnonic entanglement to the magnon modes in a dual-interferometer configuration which are prepared in a Bell state in the latter protocol. We further discuss the effect of the residual thermal excitations on the fidelity in both the protocols. Our work suggests that optomagnonic systems could become a new platform for realizing quantum teleportation and entanglement swapping where quantum superposition and entangled states of macroscopic objects (e.g., YIG spheres with diameter of hundreds of microns  \cite{optomag1,optomag2,optomag3,optomag4}) could be generated. The work may also find applications in quantum information processing and hybrid quantum networks based on magnonics.

\section*{Acknowledgments}
This research was funded by National Key Research and Development Program of China (Grant no. 2022YFA1405200) and National Natural Science Foundation of China (Grant no. 92265202).







\end{document}